# Growth, Morphology and Stability of Au in Contact with the Bi$_2$Se$_3$(0001) Surface.

M. Fanetti[a], I. Mikulska[a,†], K. Ferfolja[a], P. Moras[c], P. M. Sheverdyaeva[c], M. Panighel[d], A. Lodi-Rizzini[d], I. Píš[d,e], S. Nappini[e], M. Valant[a,b], S. Gardonio[a,*]

[a] Materials Research Laboratory, University of Nova Gorica, Vipaska 11c, Ajdovščina SI-5270, Slovenia

[b] Institute of Fundamental and Frontier Sciences, University of Electronic Science and Technology of China, Chengdu 610054, China

[c] CNR-ISM, Istituto di Struttura della Materia, S.S. 14, km 163.5, I-34149 Trieste, Italy

[d] CNR-IOM, Istituto Officina dei Materiali, S.S. 14, km 163.5, I-34149 Trieste, Italy

[e] Elettra - Sincrotrone Trieste S.C.p.A., S.S. 14, km 163.5, I-34149, Trieste, Italy

**Corresponding Author**

* E-mail: sandra.gardonio@ung.si  Phone: 00386 5 3653536

**Present Address**

[†] Diamond Light Source Ltd, Harwell Science and Innovation Campus, Didcot Oxfordshire, OX11 0DE

**Highlights**

- Au evaporated on Bi$_2$Se$_3$ at RT forms islands according to the Volmer–Weber growth mode
- upon annealing to 100° C the Au deposits are not stable and assemble into larger and thicker islands
- the Topological surface state (TSS) of Bi$_2$Se$_3$ is weakly affected by the presence of Au
- at the Au/Bi$_2$Se$_3$ interface there is a weak chemical instability that results in Bi diffusion toward the surface of Au islands and in a chemical interaction between Au and atomic Se limited at the interfacial region
- for the investigated range of Au coverages, the Au/Bi$_2$Se$_3$ heterostructure is inert towards CO and CO$_2$ exposure at low pressure (10$^{-8}$ mbar) regime

**Abstract**

We report a combined microscopy and spectroscopy study of Au deposited on the $Bi_2Se_3$(0001) single crystal surface. At room temperature Au forms islands, according to the Volmer–Weber growth mode. Upon annealing to 100° C the Au deposits are not stable and assemble into larger and thicker islands. The topological surface state of $Bi_2Se_3$ is weakly affected by the presence of Au. Contrary to other metals, such as Ag or Cr, a strong chemical instability at the Au/$Bi_2Se_3$ interface is ruled out. Core level analysis highlights Bi diffusion toward the surface of Au islands, in agreement with previous findings, while chemical interaction between Au and atomic Se is limited at the interfacial region. For the investigated range of Au coverages, the Au/$Bi_2Se_3$ heterostructure is inert towards CO and $CO_2$ exposure at low pressure ($10^{-8}$ mbar) regime.

**Introduction**

Binary bismuth chalcogenides $Bi_2Se_3$ and $Bi_2Te_3$ have been extensively studied as reference topological insulators (TIs). These materials are bulk insulators with topological surface states (TSS) crossing the Fermi level [1][2]. In contrast to conventional surface states of metals, the TSS should be extremely robust against local modifications at the surface, such as adsorbed adatoms, localized defects or changes in surface termination. This aspect makes the TIs attractive for applications in spintronics, plasmonics, quantum computing and catalysis [3][4]. In most of these applications an interface between the TI and a metal is present (e.g. for electric contacts), which makes the understanding of the structure and the stability of the metal/TI interfaces crucial, not only for future TI applications, but also for a correct theoretical modelling. In fact, depending on the near surface localization and distribution of the metal atoms on top of the TI, the electronic, magnetic, chemical and transport characteristics of the entire heterostructure can dramatically change. A theoretical model of charge transport by the TI surface states predicts that the TSS survive, provided that bonding at the metal/TI interface is weak [5]. *Ab-initio* calculations have been done to understand the electronic properties of Au, Ni, Pt, Pd and graphene layers in contact with $Bi_2Se_3$. These calculations showed that for Au and graphene the spin-momentum locking is maintained for the TSS [6], while Pd and Pt should exert a stronger interaction with the TI surface resulting in delocalization of the TSS and poor spin-momentum locking. In another study, a theory predicts that Ag and Au thin layers on $Bi_2Se_3$ have large Rashba splitting and high spin polarization of the Ag quantum wells, which give a great potential for development of spintronic devices [7]. Finally, calculations foresee also that the presence of robust TSS affect the adsorption properties of metals (bi-layer of Au, and clusters of Au, Ag, Cu, Pt, and Pd) supported on TI [8][9], in some cases resulting in the enhancement of the catalytic processes.

Despite the fundamental importance of the metal/TI interfaces and a number of theoretical studies predicting exotic phenomena, the experimental knowledge about metals on TI surfaces is surprisingly limited, especially concerning morphology and growth mode. The experimental studies reported in the literature are focused on the interactions of ultra-low coverages (almost isolated atoms) of metal atoms deposited on the TIs [10] [11] [12] [13]. A very limited number of experimental studies investigated the metal/TI systems at coverages higher than a fraction of monolayer; still, they are focused on functional aspects, such as CoNiMo alloy on $Bi_2Te_3$ nanoparticles [14] or $Pd/Bi_2Te_3$ single crystal [15], which are investigated for catalytic properties. Only a recent paper reports an experimental study of the interface chemistry between some metals on $Bi_2Se_3$ but, without a characterization of the growth mode and of the TSS evolution [16].

In this context, it is especially important to extend the studies above the diluted coverage regime and to understand what is the growth morphology of the metal on the TI surface, to what extent the metal overlayer interacts with the TI substrate, how the TSS change with the presence of the metal overlayer and what is the reactivity of the system at the different stages of the overlayer growth. Here we present a comprehensive surface sensitive characterization of Au on $Bi_2Se_3$. $Bi_2Se_3$ is considered a prototypical TI material because it has a relatively large band gap (~0.3 eV) that helps maintaining an insulating bulk and a relatively simple surface structure (one Dirac cone at the gamma point). Au was chosen because it is a very common contact metal for electronic devices and was also considered as possible co-catalyst in metal/TI based catalytic systems. In this study we have addressed the questions of the growth mode, the chemical properties at the interface and the variation of the TSS at different coverage regimes from very diluted concentrations up to tens of angstroms. Moreover, the stability and evolution of Au upon thermal treatment, have been addressed. Finally, to obtain firm experimental evidences for the chemical properties theoretically predicted for the small clusters and thin Au layer on $Bi_2Se_3$ [8][9], the Au/$Bi_2Se_3$ system was exposed to $CO_2$ and CO gas molecules to test its reactivity with respect to these two chemical species. The characterization has been done with complementary surface sensitive techniques: Scanning Electron Microscopy (SEM), Scanning Tunneling Microscopy (STM), X-ray Photoelectron Spectroscopy (XPS) and Angle Resolved Photoemission Spectroscopy (ARPES).

**Experimental methods**

Bi$_2$Se$_3$ single crystals were grown following the procedure described in *Ref.*[17]. High purity Bi (99.999%) and Se (99.999%) were mixed with an excess of selenium (Bi : Se = 2 : 3.3) in an evacuated sealed quartz ampoule and melted at 860 ºC for 24 h. Then, the ampoule was cooled down to 650 ºC with a cooling rate of 2 ºC/h and kept at this temperature for one week before cooling it down to room temperature. For the XPS and ARPES measurements the Bi$_2$Se$_3$ samples were mounted on a copper sample holder. Clean surfaces were obtained by cleaving Bi$_2$Se$_3$ *in-situ* in an ultra high vacuum (UHV) system with a pressure lower than $1.2\times10^{-9}$ mbar. Low Energy Electron Diffraction (LEED) was used to monitor the crystalline quality of the surface. Different coverage of high purity gold (99.999%) was obtained *in situ* using a well-outgassed resistively heated evaporator. The evaporator was calibrated using a quartz microbalance. During the deposition of gold, the samples were kept at room temperature. The XPS and ARPES experiments have been carried out at a VUV Photoemission beamline and BACH beamline of Elettra Synchrotron (Trieste, Italy) using a SCIENTA R4000 and SCIENTA R3000 electron analyzers, respectively. The electron analyzers were mounted 45º and 60 º with respect to the photon beam direction and the spectra were collected at normal emission. All the XPS spectra were measured using a photon energy of 650 eV with a total energy resolution (electron spectrometer and monochromator) of ~200 meV, while for the ARPES measurements the photon energy used was 20 eV with a total energy resolution of ~40 meV. All the XPS measurements were carried out at room temperature (RT) and at a base pressure lower than $1.2\times10^{-10}$ mbar. The XPS spectra were fitted with spin-orbit doublets obtained by convolution of the Doniach-Sunjic function with the Gaussian function sited on a Shirley-type background using *KolXPD* software [18]. In the fitting procedure, branching ratio, spin–orbit splitting, core-hole lifetime (Lorentzian width) and Gaussian widths were kept constant, while the intensity and energy of each doublet were considered as free parameters. In the case of Au 4*f* core level spectra, the asymmetry parameter was fixed to a non-zero value and the Gaussian width and the branching ratio were considered as free parameters. For the Bi and Se core level spectra the asymmetry parameter was set to zero. The initial stage deposition of gold on the Bi$_2$Se$_3$ surface has been studied by a scanning tunneling microscopy (STM). The STM measurements were carried out in UHV at room temperature with a commercial Omicron VT-STM. Also for STM measures, the preparation of the clean surface by cleavage and Au deposition have been performed *in situ* in the STM apparatus. STM images were acquired in constant current mode with the applied bias referred to the sample. Images were processed using the software Gwyddion [19].

Morphology and surface density of the gold clusters were characterized *ex-situ* after STM or photoemission experiment by a Scanning Electron Microscopy (SEM, JEOL JSM 7100f) with an accelerating voltage of 20 kV. The density and the fraction of covered surface were determined by a statistical analysis of the SEM images using the image processing software *ImageJ* [20].

## Results and discussion

The growth mode of Au on $Bi_2Se_3$ has been characterized by means of STM (Fig. 1) and SEM (Fig. 2 (e) – (h)). After cleaving *in-situ*, the surface has large flat areas without steps and a low defect density. The atoms are well visible at negative bias and the number of defects is low (Fig. 1 (a)). The STM analysis reveals that, at the very early stages of the growth (coverage < 2Å), Au forms symmetric dome-shaped islands, randomly distributed on the surface (Fig. 1 (b)).

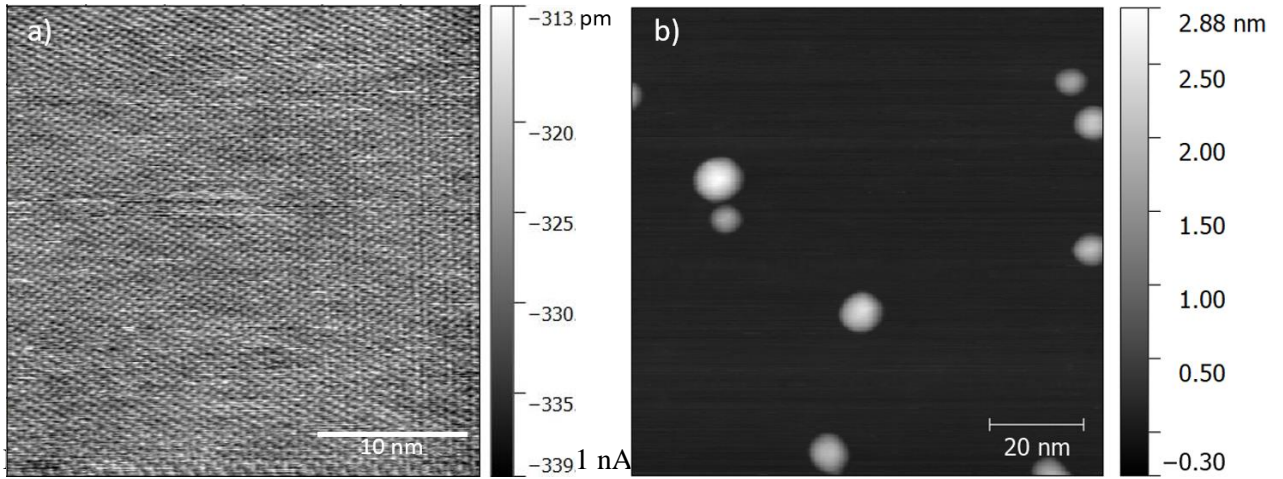

Figure 1. a) STM image (30 x 30 nm$^2$, -1.2 V, 0.1 nA) of the pristine $Bi_2Se_3$ surface after cleaving in UHV. b) STM image (100 x 100 nm$^2$, +0.5 V, 0.3 nA) of the $Bi_2Se_3$ surface after deposition of < 2 Å of Au.

Height, *h*, of the islands can be determined within a few percent accuracy from the STM images. At this coverage a statistical analysis (over 16 clusters) gave average $h = 2.14$ nm and $\sigma_h = 0.47$ nm. The width of the Au clusters determined by the STM is overestimated because of the tip shape, but can be safely evaluated to be in the range of few nanometers, as confirmed by the SEM analysis given below. Interestingly, no smaller clusters or isolated Au atoms (either as single adatoms or organized film) are observed, which indicate that Au atoms can diffuse over the surface at RT, finding a stable configuration only when they form clusters above a given threshold size, which is approximately 2 nm.

The SEM, LEED, ARPES and XPS analyses were performed in parallel on the same set of samples in order to correlate morphology, chemistry, band structure and structural properties.

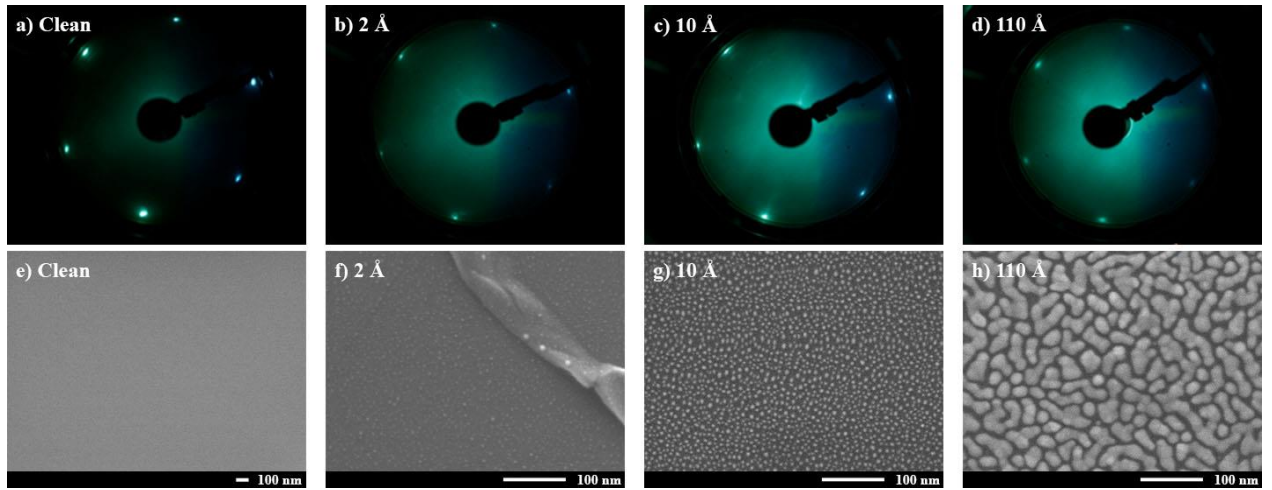

Figure 2. LEED patterns (top) and corresponding SEM images (bottom) obtained for clean $Bi_2Se_3$ surface and after different Au coverages. The LEED patterns were measured with primary electron beam energy of 44 eV.

From the analysis of the SEM images, at low coverage (2 Å, Fig. 2 (f)) the average density was found to be $6.8 \pm 1.0 \times 10^3$ islands /µm² with an average diameter of 3.8 nm ($\sigma$=0.3 nm). An increase in the Au coverage to 10 Å resulted in larger islands (average diameter 5.6 nm, $\sigma$=0.3 nm) and more densely distributed on the surface of $Bi_2Se_3$ ($9.5 \pm 1.0 \times 10^3$ islands/µm²). At this coverage the fraction of substrate area covered by the clusters is $24 \pm 2$ %. At a high Au coverage, 110 Å, the surface is predominantly covered ($63 \pm 3$ %) by coalescing islands that form elongated shapes (Fig. 2 (h)).

At the all stages of the growth, the LEED pattern shows a hexagonal symmetry (Fig. 2 (a) – (d)). The distance between the diffraction spots does not change, while the background gradually increases in intensity with the Au coverage. The lattice mismatch between Au and $Bi_2Se_3$ is of about 30% and for this reason we exclude that the observed LEED patterns correspond to pseudomorphic Au(111). The combined STM, SEM and LEED measurements revealed that Au deposited at room temperature forms three-dimensional islands starting from the early stages of the growth (Volmer-Weber mode). The observed growth resembles that of metals grown on van der Waals substrates where the metal atoms are more strongly coupled with each other than with the substrate [21]. Even for very diluted coverages, isolated or organized Au adatoms are not observed by STM, but only grouped in clusters not smaller than about 20 Å.

The electronic structure of Au on $Bi_2Se_3$ has been studied by measuring ARPES spectra along the $\overline{\Gamma} - \overline{K}$ direction in an energy region close to the Fermi level. On the data of the bare $Bi_2Se_3$ surface (Fig. 3), the conical band that corresponds to the TSS with a Dirac Point (DP) at 190 meV binding energy (BE) are clearly visible [22]. At higher BE, the TSS overlap with the bulk valence band, while near the Fermi energy the conducting band of the bulk is detected inside the surface state cone. All these features indicate an electron doping of $Bi_2Se_3$. It is known that single crystals of $Bi_2Se_3$ can exhibit a *n*-type doping behavior because of Se vacancies or anti-site defects [23]. The ARPES images were acquired 10 minutes and an hour after the cleaving of $Bi_2Se_3$. No aging effect inducing a shift of DP was observed, nor contamination adsorbed on the surface. Upon deposition of 0.3 Å of Au, there is a down-shift of the DC of about 50 meV with respect to the pristine $Bi_2Se_3$ (Fig. 3). Further Au deposition did not cause any further shift of the DP and the TSS remained well visible up to a coverage of 45 Å (Fig. 3). For even higher Au coverage the TSS are faint because of an increased contribution of the *sp* Au valence states which superimpose to that of $Bi_2Se_3$. An energy shift of the DP is consistent with previous results observed on the TIs covered with metals [10] [11][24][25], where the downward band-bending is interpreted as a result of space charge effects due to the metal adatoms. Moreover, for the entire investigated range of Au coverage there are no additional dispersing electronic states appearing close to the Fermi level as previously observed for other metals[22], and no gap opening as recently measured for Au atoms deposited at T=160 K on $Bi_2Se_3$ [26].

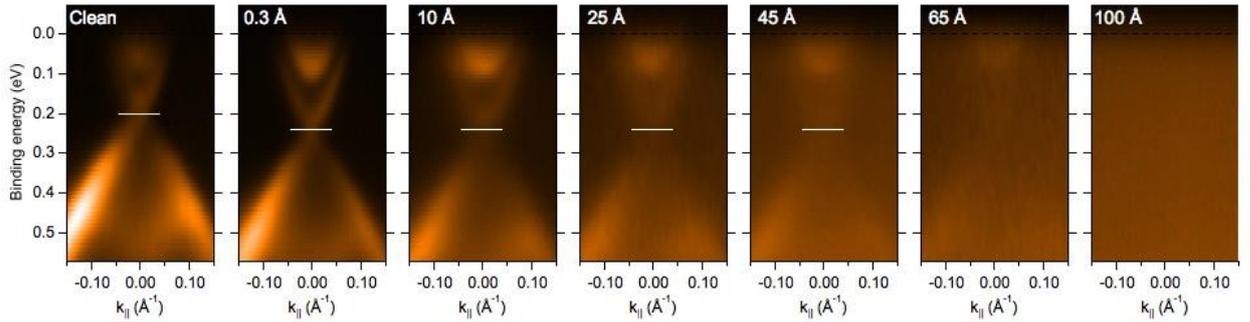

Figure 3. Electronic band structure measured near $E_F$ by means of ARPES for clean $Bi_2Se_3$ surface and after depositing different Au coverage. The white solid lines e represents the position of DP.

The chemical state characterization during the growth of Au on $Bi_2Se_3$ was performed by measuring the core level photoemission spectra of Bi 5*d*, Se 3*d* and Au 4*f*. Twelve sequential depositions were performed. Fig. 4 reports the spectra as a function of the Au coverage (results for other coverages are not shown here). At all stages of the Au deposition neither oxygen nor carbon contamination was detected. For clean $Bi_2Se_3$ the spectra of the Bi 5*d* and Se 3*d* core levels were deconvoluted with one spin-orbit doublet $Bi_1$ and $Se_1$ having the Bi $5d_{5/2}$ peak at 24.8 eV BE and the Se $3d_{5/2}$ peak at 53.3 eV BE [27][28]. For the entire range of Au coverage, the Au 4*f* spectra were fitted with one spin-orbit doublet having the Au $4f_{7/2}$ peak at 84 eV BE like metallic Au [29]. No energy shift of the core level spectra was detected within the experimental energy resolution. Going from 0.3 Å to 110 Å Au coverage the Gaussian width used to fit the Au 4*f* spectra decreased from 0.49 eV to 0.24 eV. The overall Gaussian broadening hints a statistical variation in electronic and geometric structures of the Au islands formed at low coverage.

For the Au coverage above 5 Å, the Bi 5*d* and Se 3*d* spectra are fitted with two doublets ($Bi_1$, $Bi_2$, $Se_1$ and $Se_2$). The spectral weight of $Bi_2$ with respect to $Bi_1$ increases with the Au coverage, while that of $Se_2$ remains almost unchanged with respect to $Se_1$. $Bi_2$ is at 0.75 eV lower BE with respect to $Bi_1$ and $Se_2$ is at 0.3 eV higher BE than $Se_1$. The appearance of $Bi_2$ and $Se_2$ components were observed also after depositing above 10 Å of Au in one single step. This proves that the observed chemically shifted $Bi_2$ and $Se_2$ components are not due to surface contamination or degradation occurring with time.

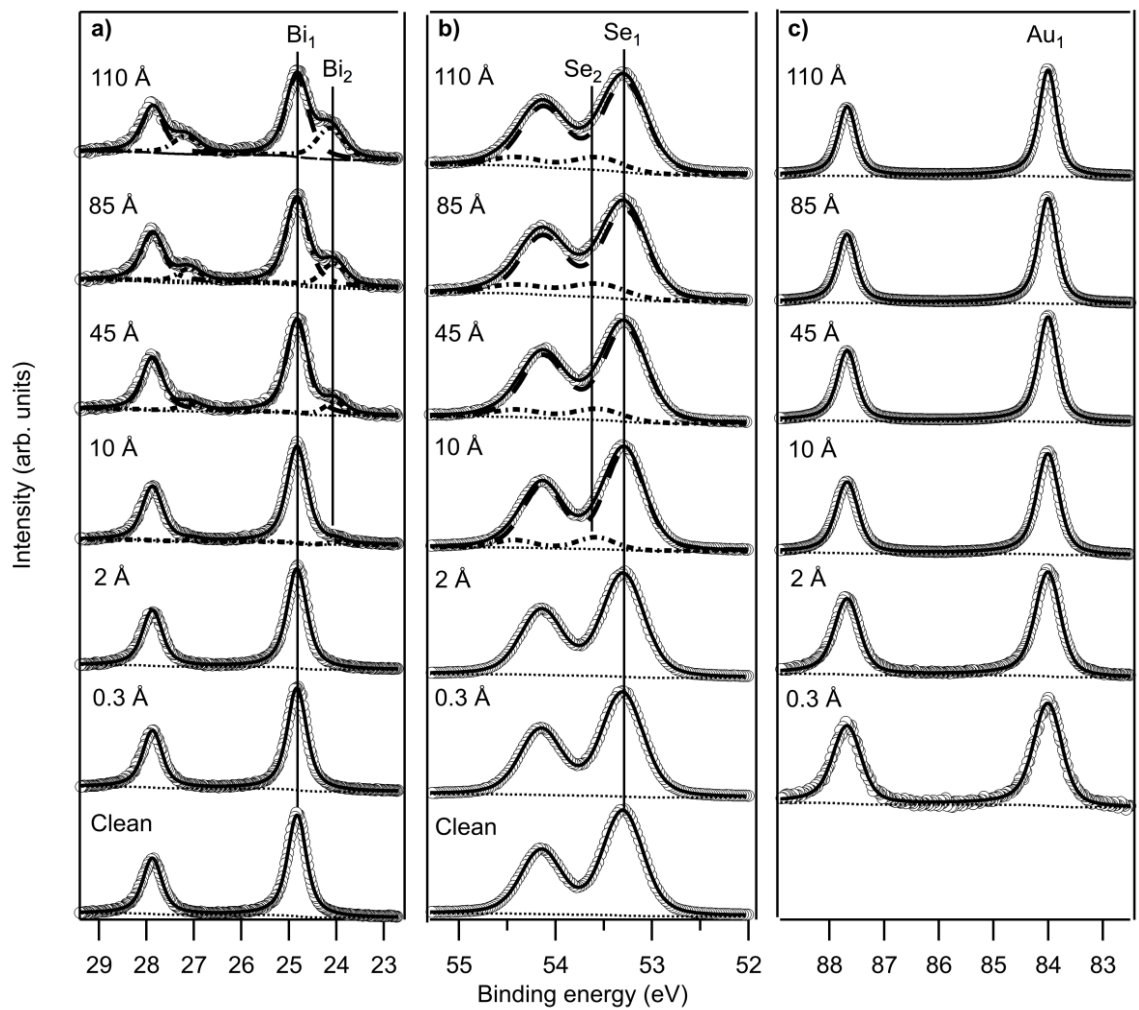

Figure 4. Core level spectra of a) Bi 5$d$, b) Se 3$d$ and c) Au 4$f$ measured on clean Bi$_2$Se$_3$ and after depositing different coverages of Au. The results of the fitting procedure are also reported. Open circles – experimental data, solid line – best fit results, long dashed line – $Bi_1$ and $Se_1$ components, dot dashed line – $Bi_2$ and $Se_2$ components and dotted line – Shirley background. The spectra are normalized to the same intensity.

To assign the $Bi_2$ and $Se_2$ components the following has to be considered. Under these experimental conditions, 95% of photoemission signal originates from the outermost ~ 24 Å of the surface [30]. This thickness is comparable with the height at a center of the Au islands, ~ 20 Å, observed here for the lowest Au coverage investigated (see above the analysis of the STM images reported in Fig.1). The photoemission signal coming from the substrate atoms in contact with the Au islands is then weakly (or not, for thicker islands occurring at higher coverage) detected in the island center, but likely detectable in the island peripheral part, where their thickness is lower. For this reason we can assume the photoemission signal as coming also from the Au/Bi$_2$Se$_3$ interface, but only in the very outermost island region.

In previous literature, new component in the spectra of Bi 5$d$ core level has been detected when Cr [31], Fe[10], Pd, Ir, Co, CoFe, Ni, Cr, NiFe , Fe and Au [16] were deposited at room temperature on Bi$_2$Se$_3$. The new component of Bi 5$d$ detected upon metals adsorption, was reported to be in the range of 0.75 eV – 1 eV lower BE with respect to the component of clean Bi$_2$Se$_3$ (Bi$^{3+}$). The BE of Bi $5d_{5/2}$ core level when bismuth is metallic is 24 eV [32], a value very close to that of *Bi$_2$*. In some cases, different BE of the new

Bi component have been found for the same metal [10] [16]. The observed component has been attributed to different possible phenomena: chemical reactivity of the metal with the substrate, out-diffusion of Bi resulting in a metallic state or more generally to a lower oxidation state of $Bi^{3+}$.

Recently, Walsh *et al.*[16] reported the transmission electron microscopy image of sharp interface formation between Au and $Bi_2Se_3$ surface, which fit with the here observed Volmer-Weber growth of Au. The same authors [16] observed the *$Bi_2$* spectral component when 15 Å of Au were deposited on $Bi_2Se_3$ and we clearly demonstrate its presence in a wide range of Au coverage (see Fig. 4). In the previous work, Bi out-diffusion was proposed as a possible explanation [16]. Here, extending the measurements to an increased number of Au thicknesses, measuring with higher energy resolution and with surface sensitivity, we demonstrate that when Au form an interface with $Bi_2Se_3$ not only Bi changes its chemical state but also Se. The *$Se_2$* core level component possesses the BE of Se atoms in contact with Au(111) surface [33]. Hence, the results indicate the occurrence of chemical interaction between Au and $Bi_2Se_3$ surface. According to microscopy measures reported in *Ref.* [16], the interaction is limited to the very interfacial region (few atomic layer), without significant interdiffusion at RT. The absence of variations in the morphology at RT as observed by SEM (contrary, for example, to Ag on $Bi_2Se_3$ [34]*)* in the timescale of weeks confirm this. The fact that the spectral weight of *$Bi_2$* with respect to *$Bi_1$* rapidly increase while *$Se_2$* does not with respect to *$Se_1$*, is ascribed to diffusion of Bi atoms into Au towards the surface of the metal islands, while Se remains localized at the interface region, which is mostly not detected with XPS, except for the very peripheral region.

Bi changes its chemical state bonding to Au and becomes metallic, while Se remains localized at the interface with a chemical state compatible with that of atoms of Se in contact with Au(111).

Thermal stability of the Au islands on $Bi_2Se_3$ was also investigated. Firstly, 1 Å of Au was deposited on $Bi_2Se_3$ at RT and subsequently annealed at 100 ºC.

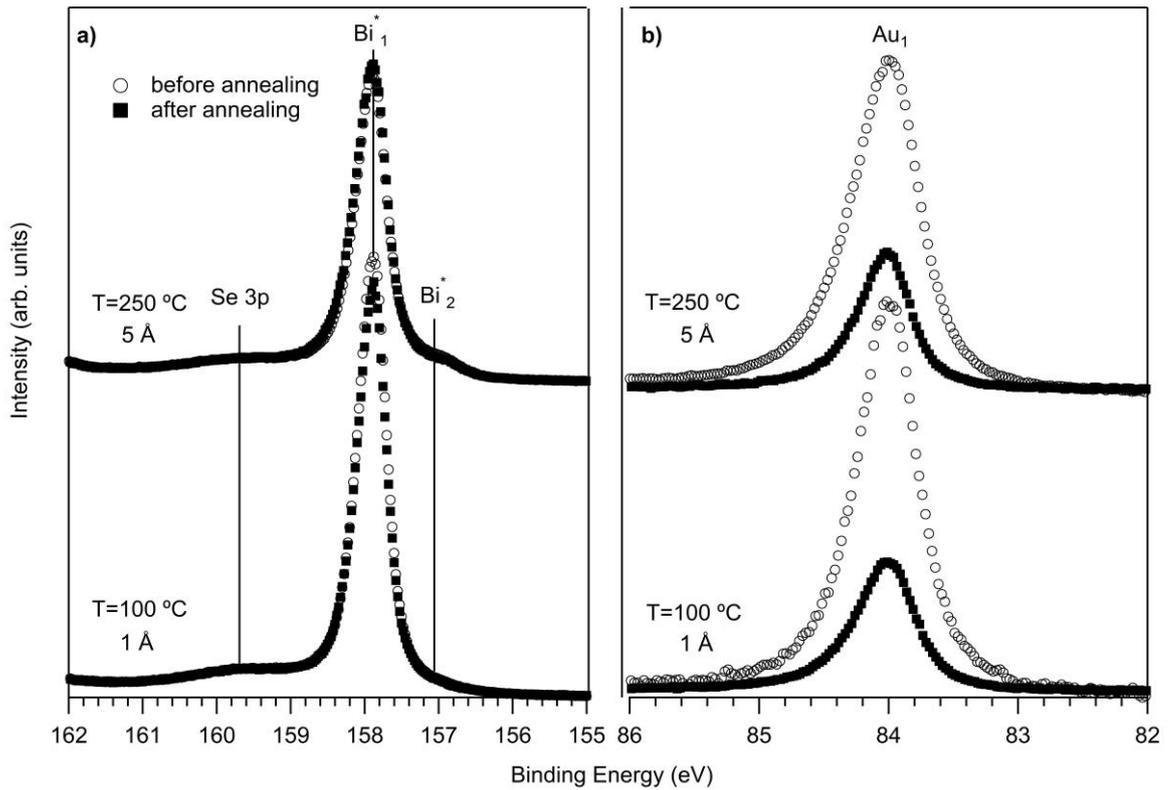

Figure 5. Core level spectra of a) Bi $4f_{7/2}$ a) and b) Au $4f_{7/2}$ after depositing 1 Å of Au on $Bi_2Se_3$ and annealing at 100 ºC (bottom spectra) and after a further deposition of 5 Å of Au and annealing at 250 ºC (top spectra). Circular markers - spectra measured immediately after deposition of gold, square markers – spectra measured after annealing.

After the first annealing at 100 ºC the Au $4f_{7/2}$ core level intensity decreases with respect to the as-deposited gold, while Bi $4f_{7/2}$ spectrum does not change its intensity and lineshape (Fig. 5 (a) and (b) bottom).

A second evaporation of Au, corresponding to a total coverage of 5 Å, was done on top of the previous one. At this second coverage, a lower BE component of Bi $4f_{7/2}$ is visible ($Bi^*_2$), in agreement with what we observed previously (Fig. 5a top, circles) Then, the sample was annealed at 250 ºC. Upon this second annealing, Bi $4f_{7/2}$ spectrum did not change its lineshape (Fig. 5 (a) top, squares), while the Au $4f_{7/2}$ spectrum decreased in its intensity (Fig. 5 (b) top).

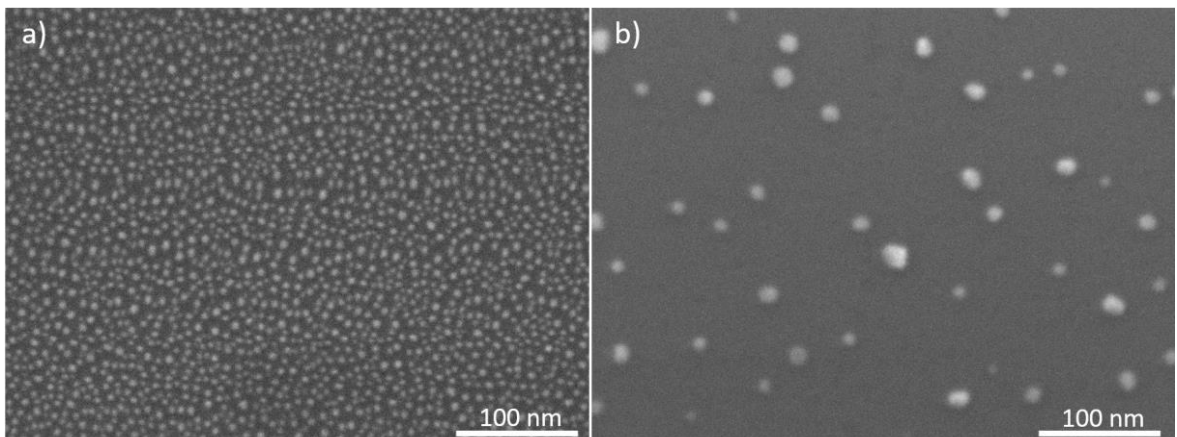

Figure 6. a) SEM image of $Bi_2Se_3$ after deposition of 10 Å of Au. b) SEM image of $Bi_2Se_3$ after deposition of 5 Å of Au and subsequent annealing at 250ºC.

In Fig. 6 (b) the SEM image of 5 Å of Au annealed at 250 ºC is shown. The density of the Au islands is much lower than that of the sample with the 10 Å Au coverage deposited at RT (Fig. 6a)). Upon the annealing the average size of the Au islands increases to ~ 15 nm. The observed morphology evolution indicates that the annealing promotes the in-plane diffusion of the Au atoms and the formation of larger and thicker gold islands. The change in the density and size of the Au clusters is the reason for the decrease in the intensity of Au 4$f$ spectra upon annealing, since the active area for the photoemission from Au is reduced.

For all the obtained Au/Bi$_2$Se$_3$ films (from 2 Å to 110 Å), the reactivity has been studied by dosing 10 L of CO and CO$_2$ at pressure of 5×10$^{-8}$ mbar at RT (1 L =1.3×10$^{-9}$ bar·s) and by measuring the XPS spectra. No signature of carbon on the surface upon dosing have been observed. The TSS and the Bi, Se and Au core levels did not change indicating the absence of the chemical reactivity of Au under these experimental conditions. Here we remark the fact that the theoretical prediction about the chemical reactivity of Au on Bi$_2$Se$_3$ was done for Au clusters of seven atoms [9], a structure that we have demonstrated to be not achievable by depositing Au on Bi$_2$Se$_3$ at RT.

**Conclusions**

Gold, deposited at RT on $Bi_2Se_3$ in a coverage range from <2 Å up to 110 Å, forms three-dimensional islands following Volmer-Weber growth mode, similarly to other systems where a substrate-overlayer interaction is weak. At more than 100 Å a network of 3D coalesced islands is observed. Our systematic investigation showed that upon the Au deposition, TSSs are stable and clearly visible with ARPES at least up to 45 Å coverage. For Au coverage above 5 Å, shifted components of Bi $5d$ and Se $3d$ core level peaks were detected in analogy to other metal/$Bi_2Se_3$ interfaces. The obtained spectroscopic results have been discussed in combination with the microscopy data. While a strong chemical instability of the metal/$Bi_2Se_3$ interface (such as for Fe, Cr, or Ag) is excluded in present case, the XPS results obtained with a surface sensitivity of about 24 Å, clearly demonstrate that a chemical interaction is present at the very interface, resulting in additional spectral components both for Se and Bi. Possible implications of this result should be evaluated especially when the system Au/$Bi_2Se_3$ is used to exploit phenomena that require chemical stability at the interface between two materials.

The Au islands are not stable upon annealing, and the effect is visible already at 100 ºC. With annealing, the coalescence of Au islands is observed, resulting in larger and less dense islands. For the entire investigated range of the gold coverage, no chemical reactivity towards CO and $CO_2$ has been detected.

The knowledge gained about the growth mode and morphology of Au on $Bi_2Se_3$ is of great importance for the design of TI based devices with optimal metal contacts, for interpretation of the observed properties (e.g. understanding of chemical shifts), but also for the appropriate modelling in theoretical investigations of metal/TI systems.

**Acknowledgements**

The work was financially supported from the Slovenian Research Agency, research core funding No. P2-0377 and P2-0379. This project has received funding from the EU-H2020 research and innovation programme under grant agreement No. 654360 having benefitted from the access provided by IOM-CNR in Trieste (Italy) within the framework of the NFFA-Europe Transnational Access Activity. We thank Tomas Skala, Federico Salvador, the staff of BACH beamline @ ELETTRA (IOM-CNR) Elena Magnano, Federica Bondino and Michele Zacchigna, and Cristina Africh of Variable and Low Temperature Scanning Tunneling Microscopy Laboratory (IOM-CNR) for the kind support.

[1]	H. Zhang, C.-X. Liu, X.-L. Qi, X. Dai, Z. Fang, S.-C. Zhang, Topological insulators in Bi2Se3, Bi2Te3 and Sb2Te3 with a single Dirac cone on the surface, Nat. Phys. 5 (2009) 438–442. doi:10.1038/nphys1270.

[2]	L. Fu, C.L. Kane, Topological insulators with inversion symmetry, Phys. Rev. B. 76 (2007) 45302. doi:10.1103/PhysRevB.76.045302.

[3]	M.Z. Hasan, C.L. Kane, Rev. Mod. Phys. 82, 3045 (2010) - Colloquium: Topological insulators, Rev. Mod. Phys. (2010). http://journals.aps.org/rmp/abstract/10.1103/RevModPhys.82.3045papers2://publication/doi/10.1103/RevModPhys.82.3045.

[4]	X.-L. Qi, S.-C. Zhang, Topological insulators and superconductors, (n.d.). doi:10.1103/RevModPhys.83.1057.

[5]	D. Culcer, E.H. Hwang, T.D. Stanescu, S. Das Sarma, Two-dimensional surface charge transport in topological insulators, Phys. Rev. B. 82 (2010) 155457. doi:10.1103/PhysRevB.82.155457.

[6]	C.D. Spataru, F. Léonard, Fermi-level pinning, charge transfer, and relaxation of spin-momentum locking at metal contacts to topological insulators, Phys. Rev. B - Condens. Matter Mater. Phys. 90 (2014). doi:10.1103/PhysRevB.90.085115.

[7]	C.-H. Chang, T.-R. Chang, H.-T. Jeng, Newtype large Rashba splitting in quantum well states induced by spin chirality in metal/topological insulator heterostructures, NPG Asia Mater. 8 (2016) e332–e332. doi:10.1038/am.2016.173.

[8]	H. Chen, W. Zhu, D. Xiao, Z. Zhang, CO oxidation facilitated by robust surface states on Au-covered topological insulators, Phys. Rev. Lett. 107 (2011). doi:10.1103/PhysRevLett.107.056804.

[9]	J. Xiao, L. Kou, C.Y. Yam, T. Frauenheim, B. Yan, Toward Rational Design of Catalysts Supported on a Topological Insulator Substrate, ACS Catal. 5 (2015) 7063–7067. doi:10.1021/acscatal.5b01966.

[10]	M.R. Scholz, J. Sánchez-Barriga, D. Marchenko, A. Varykhalov, A. Volykhov, L. V Yashina, O. Rader, Tolerance of Topological Surface States towards Magnetic Moments: Fe on ${\mathrm{Bi}}_{2}{\mathrm{Se}}_{3}$, Phys. Rev. Lett. 108 (2012) 256810. doi:10.1103/PhysRevLett.108.256810.

[11]	T. Valla, Z.-H. Pan, D. Gardner, Y.S. Lee, S. Chu, Photoemission Spectroscopy of Magnetic and Nonmagnetic Impurities on the Surface of the ${\mathrm{Bi}}_{2}{\mathrm{Se}}_{3}$ Topological Insulator, Phys. Rev. Lett. 108 (2012) 117601. doi:10.1103/PhysRevLett.108.117601.

[12]	J. Honolka, A.A. Khajetoorians, V. Sessi, T.O. Wehling, S. Stepanow, J.L. Mi, B.B. Iversen, T. Schlenk, J. Wiebe, N.B. Brookes, A.I. Lichtenstein, P. Hofmann, K. Kern, R. Wiesendanger, In-plane magnetic anisotropy of Fe atoms on Bi2Se3(111), Phys. Rev. Lett. 108 (2012) 256811. doi:10.1103/PhysRevLett.108.256811.

[13]	T. Schlenk, M. Bianchi, M. Koleini, A. Eich, O. Pietzsch, T.O. Wehling, T. Frauenheim, A. Balatsky, J.L. Mi, B.B. Iversen, J. Wiebe, A.A. Khajetoorians, P. Hofmann, R. Wiesendanger, Controllable magnetic doping of the surface state of a topological insulator, Phys. Rev. Lett. 110 (2013). doi:10.1103/PhysRevLett.110.126804.

[14]	K. Yin, Z.D. Cui, X.R. Zheng, X.J. Yang, S.L. Zhu, Z.Y. Li, Y.Q. Liang, A Bi2Te3@CoNiMo composite as a high performance bifunctional catalyst for hydrogen and oxygen evolution reactions, J. Mater. Chem. A. 3 (2015) 22770–22780. doi:10.1039/c5ta05779e.

[15]	Q.L. He, Y.H. Lai, Y. Lu, K.T. Law, I.K. Sou, Surface Reactivity Enhancement on a Pd/Bi2Te3 Heterostructure through Robust Topological Surface States, Sci. Rep. 3 (2013) 2497. doi:10.1038/srep02497.

[16]	L.A. Walsh, C.M. Smyth, A.T. Barton, Q. Wang, Z. Che, R. Yue, J. Kim, M.J. Kim, R.M. Wallace, C.L. Hinkle, Interface Chemistry of Contact Metals and Ferromagnets on the Topological Insulator Bi2Se3, J. Phys. Chem. C. 121 (2017) 23551–23563.

doi:10.1021/acs.jpcc.7b08480.

[17] M. Bianchi, R.C. Hatch, D. Guan, T. Planke, J. Mi, B.B. Iversen, P. Hofmann, The electronic structure of clean and adsorbate-covered $Bi_2Se_3$: an angle-resolved photoemission study, Semicond. Sci. Technol. 27 (2012) 124001. doi:10.1088/0268-1242/27/12/124001.

[18] Kolibrik.net, s.r.o. - Custom develompent of electronics and software, (n.d.). https://www.kolibrik.net/ (accessed June 5, 2018).

[19] D. Nečas, P. Klapetek, Gwyddion: An open-source software for SPM data analysis, Cent. Eur. J. Phys. 10 (2012) 181–188. doi:10.2478/s11534-011-0096-2.

[20] C.A. Schneider, W.S. Rasband, K.W. Eliceiri, NIH Image to ImageJ: 25 years of image analysis, Nat. Methods. 9 (2012) 671–675. doi:10.1038/nmeth.2089.

[21] a Rettenberger, P. Bruker, M. Metzler, F. Mugele, T.. Matthes, M. Böhmisch, J. Boneberg, K. Friemelt, P. Leiderer, STM investigation of the island growth of gold on WS2 and WSe2, Surf. Sci. 402–404 (1998) 409–412. doi:10.1016/S0039-6028(97)00961-8.

[22] H.-J. Noh, J. Jeong, E.-J. Cho, J. Park, J.S. Kim, I. Kim, B.-G. Park, H.-D. Kim, Controlling the evolution of two-dimensional electron gas states at a metal / $Bi_2Se_3$ interface, Phys. Rev. B. 91 (2015) 121110. doi:10.1103/PhysRevB.91.121110.

[23] J. Navrátil, J. Horák, T. Plecháček, S. Kamba, P. Lošťák, J.S. Dyck, W. Chen, C. Uher, Conduction band splitting and transport properties of Bi2Se3, J. Solid State Chem. 177 (2004) 1704–1712. doi:10.1016/j.jssc.2003.12.031.

[24] L.A. Wray, S.Y. Xu, Y. Xia, D. Hsieh, A. V. Fedorov, Y.S. Hor, R.J. Cava, A. Bansil, H. Lin, M.Z. Hasan, A topological insulator surface under strong Coulomb, magnetic and disorder perturbations, Nat. Phys. 7 (2011) 32–37. doi:10.1038/nphys1838.

[25] N. De Jong, E. Frantzeskakis, B. Zwartsenberg, Y.K. Huang, D. Wu, P. Hlawenka, J. Sańchez-Barriga, A. Varykhalov, E. Van Heumen, M.S. Golden, Angle-resolved and core-level photoemission study of interfacing the topological insulator Bi1.5Sb0.5Te1.7Se1.3 with Ag, Nb, and Fe, Phys. Rev. B - Condens. Matter Mater. Phys. 92 (2015). doi:10.1103/PhysRevB.92.075127.

[26] A. Polyakov, C. Tusche, M. Ellguth, E.D. Crozier, K. Mohseni, M.M. Otrokov, X. Zubizarreta, M.G. Vergniory, M. Geilhufe, E. V. Chulkov, A. Ernst, H.L. Meyerheim, S.S.P. Parkin, Instability of the topological surface state in Bi2Se3 upon deposition of gold, Phys. Rev. B. 95 (2017). doi:10.1103/PhysRevB.95.180202.

[27] K. Kuroda, M. Ye, E.F. Schwier, M. Nurmamat, K. Shirai, M. Nakatake, S. Ueda, K. Miyamoto, T. Okuda, H. Namatame, M. Taniguchi, Y. Ueda, A. Kimura, Experimental verification of the surface termination in the topological insulator TlBiSe2 using core-level photoelectron spectroscopy and scanning tunneling microscopy, Phys. Rev. B - Condens. Matter Mater. Phys. 88 (2013). doi:10.1103/PhysRevB.88.245308.

[28] L. Yashina, J. Sánchez-Barriga, Negligible Surface Reactivity of Topological Insulators Bi2Se3 and Bi2Te3 Towards Oxygen and Water, ACS Nano. 7 (2013) 5181. doi:10.1021/nn400908b.

[29] P.H. Citrin, G.K. Wertheim, Y. Baer, Surface-atom x-ray photoemission from clean metals: Cu, Ag, and Au, Phys. Rev. B. 27 (1983) 3160–3175. doi:10.1103/PhysRevB.27.3160.

[30] S. Tanuma, C.J. Powell, D.R. Penn, Calculations of electron inelastic mean free paths (IMFPS). IV. Evaluation of calculated IMFPs and of the predictive IMFP formula TPP-2 for electron energies between 50 and 2000 eV, Surf. Interface Anal. 20 (1993) 77–89. doi:10.1002/sia.740200112.

[31] T. Yilmaz, W. Hines, F.C. Sun, I. Pletikosić, J. Budnick, T. Valla, B. Sinkovic, Distinct effects of Cr bulk doping and surface deposition on the chemical environment and electronic structure of the topological insulator Bi2Se3, Appl. Surf. Sci. 407 (2017) 371–378. doi:10.1016/j.apsusc.2017.02.160.

[32] J.F. Moulder, W.F. Stickle, P.E. Sobol, K.D. Bomben, Standard XPS Spectra of the Elements, in: Handb. X-Ray Photoelectron Spectrosc., 1992: p. 40. doi:10.1002/sia.740030412.

[33] J. Jia, A. Bendounan, H.M.N. Kotresh, K. Chaouchi, F. Sirotti, S. Sampath, V.A. Esaulov, Selenium adsorption on Au(111) and Ag(111) surfaces: Adsorbed selenium and selenide films, J. Phys. Chem. C. 117 (2013) 9835–9842. doi:10.1021/jp4007203.

[34] K. Ferfolja, M. Valant, I. Mikulska, S. Gardonio, M. Fanetti, Chemical Instability of an Interface between Silver and $Bi_2Se_3$ Topological Insulator at Room Temperature, J. Phys. Chem. C. 122 (2018) 9980–9984. doi:10.1021/acs.jpcc.8b01543.